\documentclass[12pt]{article}
\usepackage{mathtext}         % ?? ?? ?????? ????? (? ?????)
\usepackage{amsmath,amssymb} 
\usepackage{epsfig}

\newcommand{\be}{\begin{equation}}
\newcommand{\ee}{\end{equation}}
\newcommand{\bea}{\begin{eqnarray}}
\newcommand{\eea}{\end{eqnarray}}
\newcommand{\ba}{\begin{eqnarray*}}
\newcommand{\ea}{\end{eqnarray*}}

\newcommand{\rd}{{\rm d}}

\newcommand{\Fs}[3]{\!\!\left[\begin{array}{c}#1\,;\\#2\,;\end{array}#3\right]}
\newcommand{\Fh}[2]{\,{}_#1F_#2}
\newcommand{\FmM}[2]{\Fs{#1}{#2}{ 1-\frac{m_2^2}{m_3^2} }}

\newcommand{\Fu}[2]{\Fs{#1}{#2}{1}}

\begin{document}
\begin{titlepage}

\begin{flushright}
\today  \\
\date \\
\end{flushright}

\vspace*{0.2cm}
\begin{center}
{\Large Derivation of functional equations  for Feynman integrals \\
\vspace{2mm}
 from algebraic relations
}
\\[2 cm]

\medskip

\author{
{\sc O.~V.~Tarasov}\thanks{On leave of absence from
Joint Institute for Nuclear Research,
141980 Dubna (Moscow Region), Russia.}
\\
\\
%\medskip
{\normalsize II. Institut f\"ur Theoretische Physik, Universit\"at Hamburg,}\\
{\normalsize Luruper Chaussee 149, 22761 Hamburg, Germany}\\
}
{\bf  O.V.~Tarasov}\\[0.9 cm]
%\footnote{On leave of absence from JINR,
%141980 Dubna (Moscow Region), Russian Federation.}\\[1cm]
%  \medskip 
 {
{\normalsize II. Institut f\"ur Theoretische Physik, Universit\"at Hamburg,}\\
{\normalsize Luruper Chaussee 149, 22761 Hamburg, Germany}\\
and \\
    Joint Institute for Nuclear Research,\\
      141980 Dubna, Russian Federation \\
 \medskip
     {\it E-mail}: {\tt otarasov@jinr.ru}}\\

\end{center}

\vspace*{1.0cm}

\begin{abstract}
New methods for obtaining functional equations
for Feynman integrals are presented.  
Application of these methods for finding  functional 
equations for various one- and two- loop integrals  
described in detail.
It is shown that with the aid of functional equations
Feynman integrals in general kinematics
can be expressed in terms  of  simpler integrals.  

\medskip

\medskip

\vspace{2cm}

\medskip


\end{abstract}

\end{titlepage}

%%%%%%%%%%%%%%%%%%%%%%%%%%%%%%%%%%%%%%%%%%%%%%%%%

\tableofcontents

%%%%%%%%%%%%%%%%%%%%%%%%%%%%%%%%%%%%%%%%%%%%%%%%%

\vspace*{1.0cm}

\section{Introduction}
Recently it was discovered that Feynman integrals  obey 
functional equations \cite{Tarasov:2008hw}, \cite{Tarasov:2011zz}. 
Different examples of functional equations were  presented in 
Refs.~ \cite{Tarasov:2008hw}, \cite{Kniehl:2009pv},\cite{Tarasov:2011zz}.
In  these articles only one-loop integrals were considered.

In the present paper  we propose essentially new methods 
for deriving functional equations. These methods are based on 
algebraic relations between  propagators and they are suitable 
for deriving functional equations for multi-loop integrals.
Also these methods  can be used  to derive functional equations for 
integrals with some  propagators raised to non-integer powers.

Our paper is organized as follows.
In Sec. 2. the method  proposed in Ref.~\cite{Tarasov:2008hw} is shortly reviewed.

In Sec. 3.  a method for finding algebraic relations
between products of propagators is  formulated. 
We describe in detail derivation of explicit relations for products of two, 
three  and four propagators. Also algebraic relation for  products 
of arbitrary number of proparators  is given.
These relations are used in Sec.4.  to obtain functional
equations for some one-, as well as two- loop integrals.
In particular functional equation for the massless one-loop vertex type
integral is presented. Also functional equation for the two-loop 
vertex type integral with arbitrary masses is given.

In Sec. 5. another method for obtaining functional equations
is proposed. The method is based on  finding algebraic relations 
for `deformed propagators' and  further conversion of integrals 
with `deformed  propagators' to  usual Feynman integrals by imposing 
conditions on deformation parameters. 
To perform such a conversion  the  $\alpha$- parametric representation
for both types of integrals is exploited. 
The method was  used to derive functional equation for 
the two-loop vacuum type integral with arbitrary masses.
As a by product, from this functional equation
we obtained new hypergeometric representation for the one-loop massless vertex integral.

In conclusion we formulate our vision of the future applications and 
developments of the proposed methods.

 %%%%%%%%%%%%%%%%%%%%%%%%%%%%%%%%%%%%%%%%%%%%%%%%%%%%%%%%%%%%%%%%

\section{Deriving functional equations from recurrence relations}

%%%%%%%%%%%%%%%%%%%%%%%%%%%%%%%%%%%%%%%%%%%%%%%%%%%%%%%%%%%%%%%%%

The method for deriving  functional equations proposed
in Ref.~\cite{Tarasov:2008hw} is based on the use different kind 
of recurrence relations. In particular in Refs.~\cite{Tarasov:2008hw}, 
\cite{Tarasov:2011zz}, \cite{Kniehl:2009pv},
generalized recurrence relations \cite{Tarasov:1996br} were utilized
to obtain functional equations for one-loop Feynman integrals.
In general such recurrence relations connect a combination of some number
of integrals $I_{1,n},...,I_{k,n}$ corresponding to diagrams, say, with $n$ 
lines and integrals corresponding to diagrams with fewer number of lines.
Diagrams with fewer number of lines can be obtained by 
contracting some lines in integrals with $n$ lines. 
Integrals corresponding to such diagrams depend on
fewer number of kinematical variables and masses compared
to integrals with $n$ lines. 
Such recurrence relations can be written in the following form:
%%%%%%%%%%%%%
\begin{equation}
\sum_{j} Q_j(\{m_i \},\{s_{q}\},\nu_{l},d)
 ~I_{j,~n} = \sum_{k,r<n} R_{k,r}(\{m_i\},\{s_{m}\},\nu_{l},d) 
~I_{k,~r},
\label{NconnectR}
\end{equation}
%%%%%%%%%%%%%
where $Q_j$ and $R_k$ are ratios of polynomials depending on
masses $m_i$,  scalar products $s_{r}$ of external momenta,
powers of propagators  $\nu_{l}$ and parameter of the space time dimension
$d$. At the left hand-side of Eq.~(\ref{NconnectR}) we combined
integrals  with $n$ lines and on the right hand - side
integrals with fewer number of lines.

In accordance with  the method of  Ref.~\cite{Tarasov:2008hw},
to obtain functional equation from Eq.~(\ref{NconnectR}) 
one should   eliminate terms on the left hand - side
by defining some kinematical variables from the set of equations:
%%%%%%%%%%%%
\begin{equation}
Q_j(\{m_i \},\{s_{q}\},\nu_{l},d)=0.
\label{alg_sistema}
\end{equation}
%%%%%%%%%%%%
If there is a nontrivial solution of this system and for
this solution some  $R_{k,r}(\{m_i\},\{s_{m}\},\nu_{l},d)$
are different from zero then the right-hand side of Eq.~(\ref{NconnectR}) 
will represent functional equation.

For the one-loop integrals with $n$ propagators
\begin{equation}
%(\nu_1,{\ldots} ,\nu_n)
I_n^{(d)}(\{m_j^2\},\{s_{kl}\})=
\int \frac{d^d q}{i \pi^{{d}/{2}}} \prod_{j=1}^{n}
\frac{1}{P_j^{\nu_j}},
\end{equation}
where
\begin{equation}
P_j=(q-p_j)^2-m_j^2+i\epsilon,
\end{equation}
different types of recurrence relations were given 
in Refs.~\cite{Tarasov:1996br}, \cite{Fleischer:1999hq}.
Diagram corresponding to this integral is given in Figure 1. 
\begin{figure}[ht]
\begin{center}
\includegraphics[scale=0.6]{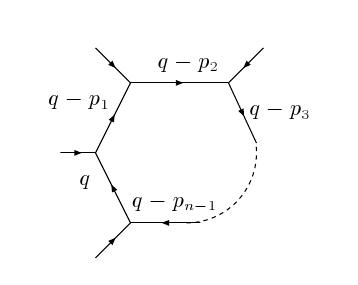}
\end{center}
\caption{One-loop  diagram with $n$ external legs }
\end{figure}
%\vspace*{0.7cm}
\noindent
In Refs.~\cite{Tarasov:1996br}, \cite{Fleischer:1999hq} the following relation
was derived: 
\begin{eqnarray}
&&(d-\sum_{i=1}^{n}\nu_i+1)G_{n-1}I^{(d+2)}_n(\{m_j^2\},\{s_{kl}\})
  -2 \Delta_n I^{(d)}_n(\{m_j^2\},\{s_{kl}\})
\nonumber \\  
&&~~~~ =
  \sum_{k=1}^n (\partial_k \Delta_n) {\bf k^-}I^{(d)}_n(\{m_j^2\},\{s_{kl}\}),
  \label{reduceDtod}
\end{eqnarray}
where the  operators   ${\bf k^{- }}$ shift index of propagators
by one unit $\nu_k \to \nu_{k } - 1$,
\begin{equation}
G_{n-1}= -2^n \left|
\begin{array}{cccc}
  (p_1-p_n)(p_1-p_n)  & (p_1-p_n)(p_2-p_n)  &\ldots & (p_1-p_n)(p_{n-1}-p_n) \\
  (p_1-p_n)(p_2-p_n)  & (p_2-p_n)(p_2-p_n)  &\ldots & (p_2-p_n)(p_{n-1}-p_n) \\
  \vdots  & \vdots  &\ddots & \vdots \\
  (p_1-p_n)(p_{n-1}-p_n)  & (p_2-p_n)(p_{n-1}-p_n)  
           &\ldots & (p_{n-1}-p_n)(p_{n-1}-p_n)
\end{array}
\right|,
\label{Gn}
\end{equation}
$$
%\Delta_n=-~\frac{(-1)^{n}}{2^n}  \left|
\Delta_n=  \left|
\begin{array}{cccc}
Y_{11}  & Y_{12}  &\ldots & Y_{1n} \\
Y_{12}  & Y_{22}  &\ldots & Y_{2n} \\
\vdots  & \vdots  &\ddots & \vdots \\
Y_{1n}  & Y_{2n}  &\ldots & Y_{nn}
\end{array}
         \right|,
$$
\begin{equation}
Y_{ij}=m_i^2+m_j^2-s_{ij},~~~~~~~~~~~~~~~s_{ij}=(p_i-p_j)^2.
\label{pij}
\end{equation}
Here $p_i, p_j$ are external momenta going through
lines $i,j$  respectively, and  $m_j$  is mass 
attributed to  $j$-th line. Gram determinant $G_{n-1}$ 
and modified Cayley determinant  $ \Delta_n$ are 
polynomials depending on scalar products and masses.

It is assumed that these scalar products are made of $d$
dimensional vectors and  $G_{n-1}$ and  $ \Delta_n$
are not subject to any restriction or condition  specific 
to some integer  values of $d$.
Eq.~(\ref{reduceDtod}) is written in the form corresponding to
Eq.~(\ref{NconnectR}). To eliminate integrals with $n$ lines 
on the left hand - side of Eq.~(\ref{reduceDtod})
the following conditions to be hold: 
\begin{equation}
G_{n-1} =0,~~~~~~~~~~~~~\Delta_n=0.
\label{sharik}
\end{equation} 

Eq.~(\ref{reduceDtod}) is valid for arbitrary kinematical variables
and masses.  Solution of Eqs.~(\ref{sharik}) 
can be easily done with respect to two kinematical variables or masses. 
Starting from $n=3$ substitution of such solutions into
Eq.~(\ref{reduceDtod}) gives nontrivial functional equations.

The method for obtaining functional equations by 
eliminating complicated integrals from recurrence relations
is quite general one.
However for multi loop integrals, depending on several 
kinematical variables, derivation of equations
like  Eq.~(\ref{reduceDtod}) is computationally challenging. 
In the next sections we will describe easier and more powerful
methods that can be used for deriving functional equations
for multi-loop integrals.

%%%%%%%%%%%%%%%%%%%%%%%%%%%%%%%%%%%%%%%%%%%%%%%%%%%%%%%%%%%%%%%%%%
%%%%%%%%%%%%%%%%%%%%%%%%%%%%%%%%%%%%%%%%%%%%%%%%%%%%%%%%%%%%%%%%%%

\section{Deriving functional equations from 
          algebraic relations \\ between propagators} 

%%%%%%%%%%%%%%%%%%%%%%%%%%%%%%%%%%%%%%%%%%%%%%%%%%%%%%%%%%%%%%%%%%
%%%%%%%%%%%%%%%%%%%%%%%%%%%%%%%%%%%%%%%%%%%%%%%%%%%%%%%%%%%%%%%%%%
Setting $\nu_j=1$  in Eq.~(\ref{reduceDtod})   and imposing conditions
(\ref{sharik}) leads to the following equation:
\begin{equation}
\sum_{k=1}^n (\partial_k \Delta_n) {\bf k^-}I^{(d)}_n=0.
\label{fe_n_points}
\end{equation}
In Eq.~(\ref{fe_n_points}) integrands of ${\bf k^-}I^{(d)}_n$
are products of $n-1$  propagators depending on different external momenta,
i.e. each term in this relation corresponds to the same function but with 
different arguments. In fact functional equations considered in 
Refs.~\cite{Tarasov:2008hw,Kniehl:2009pv,Tarasov:2011zz}
are of the same form as Eq.~(\ref{fe_n_points}).
%chosen from $n$ propagators.
%One can raise the question of whether 
The question naturally arises:
This relationship  holds
for integrals or it can be obtained  as the consequence of 
a relationship  between integrands?

By inspecting Eq.~(\ref{fe_n_points}),
one can suggest the following form of the relation
between products of propagators of  integrands:
\begin{equation}
\prod_{r=1}^{n}\frac{1}{P_r} = \frac{1}{P_{n+1}}
\sum_{r=1}^n x_r {\prod_{\substack{j=1
\\ 
j \neq r} }^n} \left(\frac{1}{P_j}\right),
\label{usual_props}
\end{equation}
where
%%%%%%%%%%%%%%
\begin{equation}
P_j=(k_1-p_j)^2 - m_j^2+i\epsilon.
\end{equation} 
%%%%%%%%%%%%%%%
In what follows we will omit $i\epsilon$ term assuming that 
all masses have such a correction.
Additionally we  assume that vectors $p_j$ are
linearly dependent, i.e. the Gram determinant for the
set  of vectors $\{p_j\}$  is equal to zero.  Such a condition
is valid  for all examples considered in 
Refs.~\cite{Tarasov:2008hw}, \cite{Tarasov:2011zz}.

Now let's consider in detail implementation of our prescription
for products of 2,3 and 4 propagators.
At $n=2$ relation (\ref{usual_props}) reads:
\begin{equation}
\frac{1}{P_1P_2}=\frac{x_1}{P_2P_3}+\frac{x_2}{P_1P_3},
\label{2prop_relation}
\end{equation}
where
\begin{equation}
P_1=(k_1-p_1)^2-m_1^2,~~~P_2=(k_1-p_2)^2-m_2^2,~~~P_3=(k_1-p_3)^2-m_3^2.
\label{P1P2P3}
\end{equation}
According to our assumption  three vectors $p_1$,$p_2$,$p_3$ are linearly 
dependent. Without loss of generality  we may assume that
\begin{equation}
p_3=y_{31}p_1 + y_{32}p_2.
\label{P3}
\end{equation}
Furthermore, we assume that $k_1$ will be integration momentum
and  scalar quantities $x_1$,$x_2$, $y_{32}$, $y_{32}$
do not depend on $k_1$.
Putting all terms in Eq.~(\ref{2prop_relation}) over a common denominator 
and then equating  to zero the  coefficients in front of various products 
of $k_1^2$, $k_1p_1$,$k_1p_2$  yields the following system 
of equations:
\begin{eqnarray}
&&y_{32} - x_2 =0,~~~y_{31} - x_1 =0,~~~ x_1+x_2=1,
\nonumber \\
&& p_1^2(x_1 - y_{31}^2) 
+ p_2^2(x_2 - y_{32}^2 ) 
+ y_{31}y_{32}(s_{12}-p_1^2-p_2^2) - m_1^2 x_1 - m_2^2x_2+ m_3^2=0.   
\label{system_for_2props}
\end{eqnarray}
Solution of this  system of equations is:
\begin{equation}
x_1=y_{31}=\lambda_2,~~~x_2=y_{32}=1-\lambda_2,
\label{y_for_2prop}
\end{equation}
where $\lambda_2$ is a root of the equation
\begin{equation}
A_2\lambda_2^2 + B_2 \lambda_2 + C_2=0,
\end{equation}
with
\begin{equation}
A_2= s_{12},~~~~B_2=m_1^2-m_2^2-s_{12},~~~~C_2=m_2^2-m_3^2.
\end{equation}
This solution can be rewritten in an explicit form:
\begin{eqnarray}
&&
x_1 = y_{31}=\frac{m_2^2-m_1^2+s_{12}}{2s_{12}}
 \pm \frac{\sqrt{\Lambda_2+4s_{12}m_3^2}}{2 s_{12}},
\nonumber \\
&&
x_2 = y_{32}=\frac{m_1^2-m_2^2+s_{12}}{2s_{12}} 
\mp \frac{\sqrt{\Lambda_2+4s_{12}m_3^2}}{2 s_{12}},
\label{solution_I2}
\end{eqnarray}
where
\begin{equation}
\Lambda_2 = s_{12}^2+m_1^4+m_2^4-2 s_{12}m_1^2-2s_{12}m_2^2-2m_1^2m_2^2.
\end{equation}

Now let's find algebraic  relation for the  products of three propagators.
At $n=3$ Eq.~(\ref{usual_props}) reads:
\begin{equation}
\frac{1}{P_1 P_2 P_3}= \frac{x_1}{P_4 P_2 P_3 }
+\frac{x_2}{P_1 P_4 P_3}+\frac{x_3}{P_1 P_2  P_4},
\label{3prop_relation}
\end{equation}
where $P_1$, $P_2$, $P_3$ are defined in Eq.(\ref{P1P2P3}) and
\begin{equation}
P_4=(k_1-p_4)^2-m_4^2.
\label{D4}
\end{equation}
In complete analogy with the previous  case we can represent 
one momentum as a combination of other ones. Without
loss of generality we may write
\begin{equation}
p_4 = y_{41}p_1+y_{42}p_2+ y_{43}p_3,
\end{equation}
where $y_{ij}$ for the time being  are arbitrary coefficients.
Putting all terms in Eq.~(\ref{3prop_relation}) over 
a common denominator and then equating to zero the coefficients
in front of various products of $(k_1^2)$, $(k_1p_1)$, $(k_1p_2)$,
$(k_1p_3)$ 
yields the following system of equations:
\begin{eqnarray} 
&&      y_{43} -  x_3=0,~~~~~
        y_{42} - x_2 =0,~~~~~
        y_{41} - x_1=0,~~~~~x_1+x_2+x_3=1,
\nonumber \\
&&
       p_1^2 ( x_1 - y_{41}^2  )
       + p_2^2 ( x_2 - y_{42}^2  )
       + p_3^2 (x_3  - y_{43}^2  )
\nonumber \\
&&
   + y_{42} y_{43}( s_{23}-p_2^2-p_3^2)
   + y_{41} y_{43}(s_{13}-p_1^2-p_3^2)
   + y_{41} y_{42} (s_{12}-p_1^2-p_2^2)
   \nonumber \\
&&~~~~~ - m^2_1 x_1- m^2_2 x_2   - m^2_3x_3 + m^2_4 =0.
\label{system_for_3props}
\end{eqnarray}
Solving these equations for $x_1$, $x_2$, $x_3$, $y_{41}$, $y_{42}$ we
have
\begin{equation}
x_1 = 1-\lambda_3-y_{43},~~~ x_2 = \lambda_3,
~~~x_3=y_{43},
~~~ y_{41} = 1-y_{43}-\lambda_3,~~~ y_{42} = \lambda_3,
\end{equation}
where $\lambda_3$ is solution of the equation
\begin{equation}
A_3\lambda_3^2 +B_3\lambda_3 +C_3=0.
\end{equation}
Here
\begin{eqnarray}
&&A_3=s_{12},\nonumber \\
&&B_3= y_{43}(s_{13}+s_{12}-s_{23})-m_1^2+m_2^2-s_{12},\nonumber \\
&&C_3= y_{43}^2s_{13} +(m_3^2-m_1^2-s_{13})y_{43}+m_1^2-m_4^2.
\end{eqnarray}

%%%%%%%%%%%%%%%%%%%%%%%%%%%%%%%%%%%%%%%%%%%%%%%%%%%%%%%%%%%%%%%
%%%%%%%%%%%%%%%%%%%%%%%%%%%%%%%%%%%%%%%%%%%%%%%%%%%%%%%%%%%%%%
Let us now turn to the derivation of algebraic relation  for the product of 
four propagators. At $n=4$ Eq.~(\ref{usual_props}) reads:
\begin{equation}
\frac{1}{P_1 P_2 P_3 P_4}= \frac{x_1}{P_5 P_2 P_3 P_4}
+\frac{x_2}{P_1 P_5 P_3 P_4}+\frac{x_3}{P_1 P_2 P_5 P_4}
+\frac{x_4}{P_1 P_2 P_3 P_5},
\label{4prop_relation}
\end{equation}
where $P_1$, $P_2$, $P_3$,$P_4$ are defined in Eqs.~(\ref{P1P2P3}),
(\ref{D4}),
\begin{equation}
P_5=(k_1-p_5)^2-m_5^2,
\end{equation}
and $p_5$ is a linear combination of vectors $p_1$,{\ldots},$p_4$,
\begin{equation}
p_5=y_{51}p_1+y_{52}p_2+y_{53}p_3+y_{54}p_4.
\end{equation}
Putting all terms in Eq.~(\ref{4prop_relation}) over a common denominator 
and then equating  to zero the coefficients in front of different products
of $k_1^2$, $k_1p_j$  yields  system of equations:
\begin{eqnarray}
&&y_{51}-x_1=0,~~~~~y_{52}-x_2=0,~~~~~y_{53}-x_3 = 0,~~~~~y_{54}-x_4 = 0,
\nonumber \\
&& x_1+x_2+x_3+x_4=1,  \nonumber \\
%&&\nonumber \\
&&~~
m_5^2-m_1^2 x_1-m_2^2x_2 -m_3^2 x_3-m_4^2x_4
+p_4^2(x_4-y_{54}^2)
\nonumber \\
&&~~
+p_3^2(x_3-y_{53}^2)+p_2^2(x_2-y_{52}^2)
+p_1^2( x_1-y_{51}^2)
+(s_{12}-p_1^2-p_2^2) y_{51}y_{52}
\nonumber \\
&&~~~~~
+(s_{13}-p_1^2-p_3^2)
y_{51}y_{53}+(s_{23}-p_2^2-p_3^2) y_{52} y_{53}
+(s_{14}-p_1^2-p_4^2)y_{54}y_{51}
\nonumber \\
&&
+~~~
(s_{24}-p_2^2-p_4^2)y_{54} y_{52}
+(s_{34}-p_3^2-p_4^2) y_{54}y_{53}=0.
\label{system_for_4props}
\end{eqnarray}
%Solution of this system reads:
Solving this system for $x_1$, $x_2$, $x_3$,$x_4$, $y_{51}$, 
$y_{54}$ we have
\begin{eqnarray}
&&x_1 = y_{51} = \lambda_4, ~~~~x_2 = y_{52},~~~~ x_3 = y_{53},  
\nonumber \\
&& x_4=y_{54}=1-x_1-x_2-x_3 = 1-y_{53}-y_{52}-\lambda_4,
\end{eqnarray}
where $\lambda_4$ is a solution of the equation
\begin{equation}
A_4 \lambda_4^2+B_4 \lambda_4+C_4=0,
\label{lambda4}
\end{equation}
with
\begin{eqnarray}
&& A_4= s_{14},
\nonumber \\
&&
B_4= (s_{24}-s_{12}+s_{14})y_{52}+(s_{34}-s_{13}+s_{14})y_{53}
+m_1^2-m_4^2-s_{14},
\nonumber \\
&&
C_4= s_{24}y_{52}^2+(s_{34}-s_{23}+s_{24})y_{52}y_{53}
+(m_2^2-m_4^2-s_{24})y_{52}+s_{34}y_{53}^2
\nonumber \\
&&~~~~~~~~~~~~~~~~~~
+(m_3^2-m_4^2-s_{34})y_{53}+m_4^2-m_5^2.
\end{eqnarray}

Eqs. (\ref{3prop_relation}), (\ref{3prop_relation}) and
(\ref{4prop_relation}) will be used in the next sections to 
derive functional equations for the propagator, vertex and
box type of integrals.
Relations between  products of five and more propagators can be 
easily derived in the same way as as it  was done for products
of two-, three- and four- propagators. 
From Eq.~(\ref{usual_props}) one can derive system of equations 
and find its solution  for arbitrary $n$. 
Multiplying both sides of Eq.~(\ref{usual_props}) by the product
of $n+1$ propagators $\prod_{j=1}^{n+1} P_j$ yields 
\begin{equation}
P_{n+1}= \sum_{r=1}^{n} P_r,
\end{equation}
or
\begin{equation}
k_1^2-2k_1p_{n+1}+p_{n+1}^2=\sum_{r=1}^n x_r(k_1^2-2k_1p_r+p_r^2 -m_r^2).
\label{ini_equ}
\end{equation}
Since we assume linear dependence of vectors $p_r$, without
loss of generality we may write:
\begin{equation}
p_{n+1}= \sum_{j=1}^n y_{n+1,j} p_j.
\label{pnp1}
\end{equation}
Substituting (\ref{pnp1}) into Eq.(\ref{ini_equ}),
collecting terms in front of $k_1^2$, $k_1p_j$ and terms
without $k_1$, equating them to zero
after some simplifications yields the following system of  $n+2$ equations:
\begin{eqnarray}
\label{Xequs}
&&x_k-y_{n+1,k}=0,~~~~~k=1,..,n,
    \\
\label{sumY}
&& \sum_{k=1}^n y_{n+1,k} = 1,  \\
\label{kwadraticY}
&&m_{n+1}^2-\sum_{k=1}^n y_{n+1,k} m_k^2 +\sum_{j=1}^{n}\sum_{l=1}^{j-1}
y_{n+1,j}y_{n+1,l}s_{lj}=0.
\end{eqnarray}
Solving Eq.~(\ref{sumY}) for one of the  $y_{ij}$ an substituting 
this  solution into Eq.~(\ref{kwadraticY}) gives quadratic equation
for the remaining $y_{ij}$. This quadratic equation can be solved
with respect to  one of the parameters $y_{ij}$. Thus the solution of  the system
of equations (\ref{Xequs}), (\ref{sumY}), (\ref{kwadraticY}) will depend on
$n-2$ arbitrary parameters $y_{ij}$ and one arbitrary mass $m_n$.

It is interesting to note that for any $n$, functional equations
for integrals with all masses equal to zero and functional 
equations for integrals with  all masses  equal are the same.
In case of equal masses,  two mass dependent terms in 
Eq.~(\ref{kwadraticY}) cancel each other due to Eq.~(\ref{sumY}).
In both cases systems of equations for $x_i$, $y_{jk}$ are the
same and therefore arguments of integrals are the same.

Eq.~(\ref{2prop_relation})  is analogous to the equation for splitting 
propagators presented  in Ref.~\cite{'tHooft:1978xw}.   Eq.~(\ref{3prop_relation}) is a 
generalization of  Eq.~(\ref{2prop_relation}). Indeed, setting 
$y_{43}=x_3=0$, canceling  common factor $P_3$ on both sides of 
Eq.~(\ref{3prop_relation}) yields relation similar to 
(\ref{2prop_relation}). In turn Eq.~(\ref{4prop_relation})
is a generalization of (\ref{3prop_relation}).

%%%%%%%%%%%%%%%%%%%%%%%%%%%%%%%%%%%%%%%%%%%%%%%%%%%
\subsection{Prototypes of functional equations}
%%%%%%%%%%%%%%%%%%%%%%%%%%%%%%%%%%%%%%%%%%%%%%%%%%%

Multiplying algebraic relations (\ref{2prop_relation}),(\ref{3prop_relation}),
(\ref{4prop_relation})   by products of any number of propagators 
raised to arbitrary powers $\nu_j$ 
\begin{equation}
\prod_{j=n_0}^{N} \frac{1}{[(k_1-p_j)-m_j^2]^{\nu_j}}
\end{equation}
and integrating with respect to $k_1$ we get a functional equation
for  one-loop integrals.
Eqs.~ (\ref{2prop_relation}), (\ref{3prop_relation}),
(\ref{4prop_relation})  also can be used to derive functional 
equations for integrals with any number of loops. 
Multiplying algebraic relations for propagators  by function
corresponding to Feynman integral depending on momentum $k_1$ and any
number of external momenta and then integrating with respect to
$k_1$ will produce functional equations.  
 Just for demonstrational purposes we present graphically in Figure 2
 functional equation based on $n$  propagator relation.
\begin{figure}[ht]
\begin{center}
\includegraphics[scale=0.5]{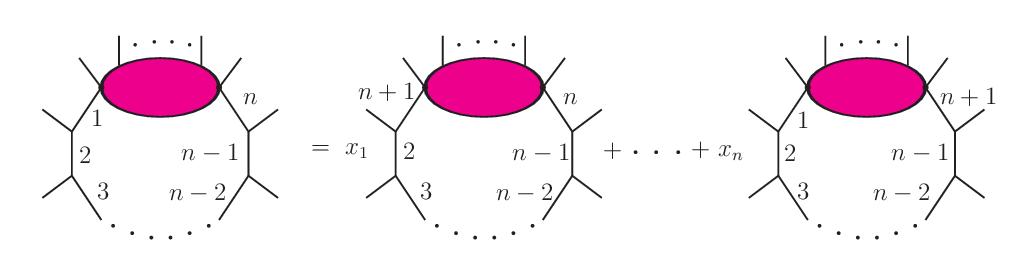}
\end{center}
\caption{$n$ - propagator  functional equation}
\end{figure}
The blob on this picture correspond to either
product of propagators raised to arbitrary powers
or to an integral with any number of loops and external legs. 
One of the external momenta of this multi loop integral
should be $k_1$.

%%%%%%%%%%%%%%%%%%%%%%%%%%%%%%%%%%%%%%%%%%%%%%%%%%%%%%%%%%%%%%%%%%
\section{Some examples of functional equations}

In this section several particular examples of functional
equations resulting from algebraic relations for 
products of propagators will be considered.
%%%%%%%%%%%%%%%%%%%%%%%%%%%%%%%%%%%%%%%%%%%%%%%%%%%%%%%%%%%%%%%%%%
\subsection{Functional equation for the one-loop 
propagator  type integral }
%%%%%%%%%%%%%%%%%%%%%%%%%%%%%%%%%%%%%%%%%%%%%%%%%%%%%%%%%%%%%%%%%%

%In order to  answer  this question 
First, we   consider the simplest case,   namely,   
functional equation for the integral  $I^{(d)}_{2}$:
\begin{equation}
I_2^{(d)}(m_j^2,m_k^2;~s_{jk})=
\int \frac{d^d k_1}{i \pi^{{d}/{2}}}
\frac{1}{[(k_1-p_j)^2-m_j^2]
         [(k_1-p_k)^2-m_k^2]}.
\end{equation}
Integrating both sides of Eq.~(\ref{2prop_relation}) with respect to
$k_1$, we get:
\begin{equation}
 I_2^{(d)}(m_1^2,m_2^2,s_{12})=
 x_1I_2^{(d)}(m_2^2,m_3^2,s_{23})
+x_2 I_2^{(d)}(m_1^2,m_3^2,s_{13}).
\label{prop_fe}
\end{equation} 
The arguments $s_{13}$, $s_{23}$ of integrals on the right hand - side depend on
$y_{31}$, $y_{32}$
\begin{eqnarray}
s_{13}=(p_1-p_3)^2=(y_{31}-1)^2 p_1^2+2 y_{32} (y_{31}-1)p_1p_2+y_{32}^2
p_2^2,
\nonumber \\
s_{23}=(p_2-p_3)^2=p_1^2y_{31}^2+2(y_{32}-1)y_{31}p_1p_2+(y_{32}-1)^2p_2^2.
\label{p13_p23}
\end{eqnarray}
Substituting  solution for  $y_{ij}$ from Eq.~(\ref{y_for_2prop}) 
into Eq.~(\ref{p13_p23}) yields:
%From (\ref{P3}), (\ref{solution_I2}) it follows that
\begin{eqnarray}
&&s_{13} = \frac{m_2^4-2 m_2^2 m_1^2-2 m_2^2s_{12}
+m_1^4+s_{12}^2+2 s_{12}m_3^2}{2 s_{12}}
\nonumber \\
&&~~~~~~~~~~~~~~~~~~~~~~~~~~~~~~~
\pm \frac{m_1^2-m_2^2+s_{12}}{2s_{12}} \sqrt{\Lambda_2+4s_{12}m_3^2}
\nonumber 
\\
&&
s_{23} =
 \frac{s_{12}^2-2m_1^2 s_{12}+m_1^4 +m_2^4-2m_1^2 m_2^2
+2s_{12}m_3^2}
{2s_{12}}
\nonumber \\
&&~~~~~~~~~~~~~~~~~~~~~~~~~~~~~~~
\pm \frac{m_1^2-m_2^2-s_{12}}{2s_{12}}\sqrt{\Lambda_2+4s_{12}m_3^2}. 
\label{S13_S23}
\end{eqnarray}
%It should be noticed that 
In this equation  $m_3^2$ is an arbitrary parameter and can be taken at will.
Functional equation (\ref{prop_fe}) is in agreement with the result
presented in Refs.~\cite{Tarasov:2008hw},\cite{Tarasov:2011zz}.
%%%%%%%%%%%%%%%%%%%%%%%%%%%%%%%%%%%%%%%%%%%%%%%%%%%%%%%%%%%%%%%%%%
\subsection{Functional equations for the one-loop 
 vertex  type integral }
%%%%%%%%%%%%%%%%%%%%%%%%%%%%%%%%%%%%%%%%%%%%%%%%%%%%%%%%%%%%%%%%%%

Functional equations for the vertex type integral
\begin{eqnarray}
&&I_3^{(d)}(m_1^2,m_2^2,m_3^2,s_{23},s_{13},s_{12} )
\nonumber \\
&&
~~~~~~~~~~=
\int \frac{d^d k_1}{i \pi^{{d}/{2}}}
\frac{1}{[(k_1-p_1)^2-m_1^2]
         [(k_1-p_2)^2-m_2^2]
	 [(k_1-p_3)^2-m_3^2]
	 },
\label{I3definition}	 
\end{eqnarray}
%From Eq.(\ref{2prop_relation}) one can obtain functional
%equations for one-loop $n$-point ($n>2$) integrals.
%Functional equation for $I_3^{(d)}$  
can be obtained from Eq.~(\ref{2prop_relation}) as well as from
Eq.~(\ref{3prop_relation}).
Multiplying Eq.~(\ref{2prop_relation}) 
with the factor ${1}/{P_4}$ where
\begin{equation}
P_4={(k_1-p_4)^2-m_4^2},
\end{equation}
and integrating over $k_1$ leads to the equation: 
\begin{equation}
\int \frac{d^dk_1}{P_1 P_2 P_4}
=x_1\int \frac{d^dk_1}{P_2 P_3 P_4}+x_2\int \frac{d^dk_1}{P_1 P_3 P_4}.
\label{fe_for_vertex}
\end{equation}
This equation  in terms of integrals $I_3^{(d)}$ reads
\begin{eqnarray}
&&I_3^{(d)}(m_1^2,m_2^2,m_4^2;s_{24},s_{14},s_{12})
\nonumber \\
&&
~~~~~~~=
x_1I_3^{(d)}(m_2^2,m_3^2,m_4^2;s_{34},s_{24},s_{23})
+x_2 I_3^{(d)}(m_1^2,m_3^2,m_4^2;s_{34},s_{14},s_{13}).
\label{feI3massiv}
\end{eqnarray}
Two more functional equations can be obtained from 
Eq.~(\ref{fe_for_vertex}) by symmetric permutations 
$(m_1,p_1) \leftrightarrow (m_4,p_4)$ and
$(m_2,p_2) \leftrightarrow (m_4,p_4)$.

Another  functional equation for the vertex type integral
can be obtained by 
integrating  Eq.~(\ref{3prop_relation}) with respect to $k_1$:
\begin{eqnarray}
&&I_3^{(d)}(m_1^2,m_2^2,m_3^2,s_{23},s_{13},s_{12})
=(1-\lambda_3-y_{43})
I_3^{(d)}(m^2_{4},m_2^2,m_3^2,s_{23},s_{34},
{s}_{24})
\nonumber \\
&&\nonumber \\
&&~~~~~+\lambda_3 
I_3^{(d)}(m_1^2,m_{4}^2,m_3^2,{s}_{34},
{s}_{13}, s_{14})
+y_{43} I_3^{(d)}(m_1^2,m_2^2,m_{4}^2,s_{24},  
{s}_{14},{s}_{13}),
\label{feI3massive}
\end{eqnarray}
where 
\begin{eqnarray}
&&{s}_{14}=(m^2_1-m^2_2+s_{12})\lambda_3+(m^2_1-m^2_3+s_{13})y_{43}
+m^2_4-m^2_1,
\nonumber \\
&&{s}_{24}=(m^2_1-s_{12}-m^2_2)\lambda_3
+(m_1^2-m_3^2-s_{12}+s_{23})y_{43}+m^2_4-m^2_1+s_{12},
\nonumber \\
&&{s}_{34}=(m^2_1-s_{13}+s_{23}-m^2_2)\lambda_3
+(m^2_1-m^2_3-s_{13})y_{43}+m^2_4-m^2_1+s_{13}.
\end{eqnarray}
There is an essential difference between functional equation Eq.~(\ref{feI3massiv})
obtained from Eq.~(\ref{2prop_relation}) and functional equation (\ref{feI3massive})
derived from Eq.~(\ref{3prop_relation}).
For example, at $m_1^2=m_2^2=m_3^2=m_4^2=0$,  Eq.~(\ref{feI3massiv}) 
becomes trivial while  from Eq.~(\ref{feI3massive}) for the integral 
\begin{equation}
I_3^{(d)}(0,0,0,s_{23},s_{13},s_{12})\equiv 
I_3^{(d)}(s_{23},s_{13},s_{12})
\end{equation}
we obtain nontrivial  functional equation: 
\begin{eqnarray}
&&I^{(d)}_3(s_{23},s_{13},s_{12})
\nonumber \\
&&= (1-\lambda_3-y_{43})
   I^{(d)}_3(
   s_{12}(1-\lambda_3-y_{43})+s_{23}y_{43},
   s_{23}, 
   \lambda_3 (s_{23}-s_{13})+s_{13}(1-y_{43}) )
\nonumber \\
&&
   +\lambda_3
   I^{(d)}_3(s_{13}y_{43}+\lambda_3 s_{12},
   \lambda_3  (s_{23}-s_{13}) +s_{13}(1-y_{43}),s_{13})
\nonumber \\
&&
   +y_{43}
   I^{(d)}_3(s_{12},s_{12}(1-\lambda_3-y_{43})
   +s_{23}y_{43},s_{13}y_{43}+\lambda_3 s_{12}),
\label{FE_triangle}
\end{eqnarray}
where $\lambda_3$ is a root of the quadratic equation
\begin{equation}
s_{12} \lambda_3^2
-(s_{12}-s_{12}y_{43}-s_{13}y_{43}+s_{23}y_{43})\lambda_3
+y_{43}^2s_{13}-y_{43}s_{13}=0.
\label{lambda3_zero_masses}
\end{equation}

If one argument of  $I_3^{(d)}(s_{23},s_{13},s_{12})$ is zero then by
applying functional equation (\ref{FE_triangle}) such an integral can
be expressed in terms of integrals $I_3^{(d)}$
with two arguments equal to zero. For example,
at $s_{23}=0$ and $y_{43}=s_{12}/(s_{12}-s_{13})$
the  relation (\ref{FE_triangle}) becomes: 
\begin{equation}
I_3^{(d)}(0,s_{13},s_{12}) =
\frac{s_{12}}{s_{12}-s_{13}} I_3^{(d)}(s_{12},0,0)
-\frac{s_{13}}{s_{12}-s_{13}} I_3^{(d)}(0,0,s_{13}).
\label{1zero_2zeros}
\end{equation}
This is a typical example how functional equations can be used 
to simplify evaluation of  an integral by reducing it to a combination 
of integrals with fewer number of arguments.

At $m_1^2=m_2^2=m_3^2=m_4^2=m^2$, similar to the previous case,
Eq.(\ref{feI3massiv}) 
degenerate while  from Eq.(\ref{feI3massive}) for the integral 
\begin{equation}
I_3^{(d)}(m^2,m^2,m^2,s_{23},s_{13},s_{12})\equiv 
I_3^{(d)}(m^2;s_{23},s_{13},s_{12})
\end{equation}
we obtain nontrivial  functional equation: 
\begin{eqnarray}
&&I^{(d)}_3(m^2;s_{23},s_{13},s_{12})
\nonumber \\
&&= (1-\lambda_3-y_{43})
   I^{(d)}_3(m^2;
   s_{12}(1-\lambda_3-y_{43})+s_{23}y_{43},
   s_{23}, 
   \lambda_3 (s_{23}-s_{13})+s_{13}(1-y_{43}) )
\nonumber \\
&&
   +\lambda_3
   I^{(d)}_3(m^2;s_{13}y_{43}+\lambda_3 s_{12},
   \lambda_3  (s_{23}-s_{13}) +s_{13}(1-y_{43}),s_{13})
\nonumber \\
&&
   +y_{43}
   I^{(d)}_3(m^2;s_{12},s_{12}(1-\lambda_3-y_{43})
   +s_{23}y_{43},s_{13}y_{43}+\lambda_3 s_{12}),
\label{FE_triangle_eqm}
\end{eqnarray}
where $\lambda_3$ is a root of the quadratic equation
\begin{equation}
s_{12} \lambda_3^2
-(s_{12}-s_{12}y_{43}-s_{13}y_{43}+s_{23}y_{43})\lambda_3
+y_{43}^2s_{13}-y_{43}s_{13}=0.
\label{L3_eqm}
\end{equation}
Eqs.~(\ref{FE_triangle_eqm}), (\ref{L3_eqm}) are identical to 
Eqs.~(\ref{FE_triangle}),(\ref{lambda3_zero_masses}) respectively and therefore
functional equation for the integral with massless propagators and
functional equation for the integral with all masses equal 
are the same.
Eq.~(\ref{FE_triangle_eqm})  at 
$s_{23}=0$ and $y_{43}=s_{12}/(s_{12}-s_{13})$ leads to the relation
similar to (\ref{1zero_2zeros}):
\begin{equation}
I_3^{(d)}(m^2;0,s_{13},s_{12}) =
\frac{s_{12}}{s_{12}-s_{13}} I_3^{(d)}(m^2;s_{12},0,0)
-\frac{s_{13}}{s_{12}-s_{13}} I_3^{(d)}(m^2;0,0,s_{13}).
\label{1zer2zers_eqm}
\end{equation}
This is not surprising because coefficients of the 
Eq.~(\ref{FE_triangle_eqm}) are mass independent and
in the integrand $m^2$ and $i\epsilon$ appear in the covariant
combination $m^2-i\epsilon$. For this reason the similarity
of functional equations for massless integrals and integrals 
with all masses equal take place for integrals with more
external legs and more loops.
%%%%%%%%%%%%%%%%%%%%%%%%%%%%%%%%%%%%%%%%%%%%%%%%%%%%%%%%%%%%%%
%%%%%%%%%%%%%%%%%%%%%%%%%%%%%%%%%%%%%%%%%%%%%%%%%%%%%%%%%%%%%%

\subsection{Functional equations for one-loop box type integrals}

%%%%%%%%%%%%%%%%%%%%%%%%%%%%%%%%%%%%%%%%%%%%%%%%%%%%%%%%%%%%%%
%%%%%%%%%%%%%%%%%%%%%%%%%%%%%%%%%%%%%%%%%%%%%%%%%%%%%%%%%%%%%%

Functional equations for the box type integrals can be obtained
by multiplying  relation (\ref{2prop_relation}) by two propagators,
or by multiplying  relation (\ref{3prop_relation}) by one propagator
and then integrating over momentum $k_1$. Yet
another relation can be obtained just by integrating Eq.~(\ref{4prop_relation})
over momentum $k_1$:
%When all $m_j^2=0$ Eq.(\ref{feI3massiv}) becomes trivial.
\begin{eqnarray}
&&I_4^{(d)}(m_1^2,m_2^2,m_3^2,m_4^2;
s_{12},s_{23},s_{34},s_{14},s_{24},s_{13})
\nonumber \\
&&~~~~~~~~~~~~~~~~~~~~=\lambda_4
I_4^{(d)}(m_5^2,m_2^2,m_3^2,m_4^2;
s_{25},s_{23},s_{34},s_{45},s_{24},s_{35})
\nonumber \\
&&~~~~~~~~~~~~~~~~~~~~+y_{52}I_4^{(d)}(m_1^2,m_5^2,m_3^2,m_4^2;
s_{15},s_{35},s_{34},s_{14},s_{45},s_{13})
\nonumber \\
&&~~~~~~~~~~~~~~~~~~~~+y_{53}I_4^{(d)}(m_1^2,m_2^2,m_5^2,m_4^2;
s_{12},s_{25},s_{45},s_{14},s_{24},s_{15})
\nonumber \\
&&+(1-y_{52}-y_{53}-\lambda_4)
I_4^{(d)}(m_1^2,m_2^2,m_3^2,m_5^2;
s_{12},s_{23},s_{35},s_{15},s_{25},s_{13}).
\label{box_func_equ}
\end{eqnarray}
Here $\lambda_4$ is defined in Eq.~(\ref{lambda4}) and $m_5$,$y_{52}$,
$y_{53}$ are arbitrary parameters and 
\begin{eqnarray}
&&s_{15} = \lambda_4 (m_4^2- s_{14}  - m_1^2 )
       + y_{53} (m_4^2 - m_3^2  - s_{14} + s_{13}  )
       + y_{52}  (m_4^2 - m_2^2 - s_{14} + s_{12}  )
       \nonumber \\
&&~~~~~~~~~~~~~~~~~~~~~~~~~~~~~~~~~~~~~~~~~~~~~~~~~~       
       + s_{14} + m_5^2 - m_4^2,
\nonumber \\
&&s_{25} =
       \lambda_4 (m_4^2 - m_1^2 - s_{24} + s_{12}  )
       + y_{53}  (m_4^2 - m_3^2  - s_{24} + s_{23}  )
       + y_{52}  (m_4^2 - m_2^2 - s_{24}  )
       \nonumber \\
&&~~~~~~~~~~~~~~~~~~~~~~~~~~~~~~~~~~~~~~~~~~~~~~~~~~       
              + s_{24} + m_5^2 - m_4^2,
\nonumber \\
&&s_{35} =
        \lambda_4 (m_4^2 - m_1^2  - s_{34} + s_{13}  )
       + y_{53}  (m_4^2 - m_3^2- s_{34}  )
       + y_{52} (m_4^2 - m_2^2- s_{34} + s_{23}  )
       \nonumber \\
&&~~~~~~~~~~~~~~~~~~~~~~~~~~~~~~~~~~~~~~~~~~~~~~~~~~       
              + s_{34} + m_5^2 - m_4^2,
\nonumber \\
&&s_{45} =
        \lambda_4 ( s_{14} + m_4^2 - m_1^2 )
       + y_{53} ( s_{34} + m_4^2 - m_3^2 )
       + y_{52}  ( s_{24} + m_4^2 - m_2^2 )
       \nonumber \\
&&~~~~~~~~~~~~~~~~~~~~~~~~~~~~~~~~~~~~~~~~~~~~~~~~~~       
              + m_5^2 - m_4^2.
\end{eqnarray}
Arbitrary parameters in this functional equation
can be chosen from the requirement of simplicity of
evaluation of integrals on the right hand - side of
Eq.~(\ref{box_func_equ}) or from some other requirements.
For example, one can choose these parameters by 
transforming arguments to a certain kinematical region
needed for analytic continuation of the original integral.

%%%%%%%%%%%%%%%%%%%%%%%%%%%%%%%%%%%%%%%%%%%%%%%%%%%%%%%%%

\subsection{Functional equation for two-loop vertex type integral}

%%%%%%%%%%%%%%%%%%%%%%%%%%%%%%%%%%%%%%%%%%%%%%%%%%%%%%%%%

The method described in the previous  section 
can be  applied to multi loop integrals. Consider, for example, integral
corresponding to the diagram given in Figure 3.
\begin{figure}[ht]
\begin{center}
\includegraphics[scale=0.5]{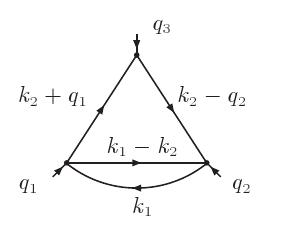}
\end{center}
\caption{ Diagram corresponding to $R(m_1^2,m_2^2,m_3^2,m_4^2;~q_1^2,q_2^2,q_3^2)$ }
\end{figure}
If we multiply  Eq.~(\ref{2prop_relation}) by the one-loop integral
depending on $k_1$
\begin{equation}
\int \frac{d^dk_2}{[k_2^2-m_4^2] [( k_1-k_2)^2-m_5^2]}
\end{equation}
and integrate with respect to  momentum $k_1$ then we obtain functional equation 
\begin{eqnarray}
&&R(m_1^2,m_2^2,m_3^2,m_4^2;~q_1^2,q_2^2,q_3^2)
\nonumber\\
&&~~~~~~~~~~~~~= 
\alpha R(0,m_2^2,m_3^2,m_4^2;~Q^2,q_2^2,(m_2^2-m_1^2+q_3^2)\alpha-m_2^2)
\nonumber
\\
&&
~~~~~~~~~~~~~+(1-\alpha)
R(m_1^2,0,m_3^2,m_4^2;~q_1^2,Q^2,(m_2^2-m_1^2-q_3^2)\alpha+q_3^2-m_2^2),
\label{fe_for_R}
\end{eqnarray}
%%%%%%%%%%%%%%%%%%%%%%%%%%%%%%%%%%%%%%%%%%%%%%%%%%%%%%%%%
where
\begin{eqnarray}
&&R(m_1^2,m_2^2,m_3^2,m_4^2;~q_1^2,q_2^2,q_3^2)
\nonumber
\\
&&=\int \int
\frac{d^dk_1d^dk_2}{(i\pi^{d/2})^2}\frac{1}
{[(k_2+q_1)^2-m_1^2][(k_2-q_2)^2-m_2^2]
[k_1^2-m_3^2][(k_1-k_2)^2-m_4^2]},
\label{Rdefinition}
\end{eqnarray}
\begin{eqnarray}
&& Q^2=(q_1^2-q_2^2-m_1^2+m_2^2)\alpha +q_2^2-m_2^2, \\
&& \alpha=\frac{q_3^2-m_1^2+m_2^2 \pm \sqrt{\Delta}}{2q_3^2},\\
&& \Delta = q_3^4+m_1^4+m_2^4-2q_3^2m_1^2-2q_3^2m_2^2-2m_1^2m_2^2.
\end{eqnarray}
Integrals of this type arise, for example, in calculations of two-loop 
radiative corrections in the electroweak theory.
Instead of the integral $R(m_1^2,m_2^2,m_3^2,m_4^2;~q_1^2,q_2^2,q_3^2)$ 
one can consider  derivative of $R$  with respect to  $m_3^2$ which is
UV finite:
\begin{eqnarray}
&&\frac{\partial R}{\partial m_3^2} \equiv R_3(m_1^2,m_2^2,m_3^2,m_4^2;~q_1^2,q_2^2,q_3^2)
\nonumber
\\
&&=\int \int
\frac{d^dk_1d^dk_2}{(i\pi^{d/2})^2}\frac{1}
{[(k_2+q_1)^2-m_1^2][(k_2-q_2)^2-m_2^2]
[k_1^2-m_3^2]^2[(k_1-k_2)^2-m_4^2]}.
\label{R3definition}
\end{eqnarray}
Integral $R_3$ satisfy the following functional equations: 
\begin{eqnarray}
%\label{integralR3}
&&R_3(m_1^2,m_2^2,m_3^2,m_4^2;~q_1^2,q_2^2,q_3^2)
\nonumber\\
&&~~~~~~~~~~~~~= 
\alpha R_3(0,m_2^2,m_3^2,m_4^2;~Q^2,q_2^2,(m_2^2-m_1^2+q_3^2)\alpha-m_2^2)
\nonumber
\\
&&
~~~~~~~~~~~~~+(1-\alpha)
R_3(m_1^2,0,m_3^2,m_4^2;~q_1^2,Q^2,(m_2^2-m_1^2-q_3^2)\alpha+q_3^2-m_2^2).
\label{fe3}
\end{eqnarray}
This relation can be used for computing basis integral
arising in calculation of two-loop radiative correction to the ortho
-positronium lifetime. In particular one
of these basis integrals corresponds to kinematics
$m_1^2=m_2^2=m_3^2=m_4^2=m^2$, $q_1^2=q_2^2=m^2$, $q_3^2=4m^2$.
In this case relation (\ref{fe3}) reads
\begin{equation}
R_3(m^2,m^2,m^2,m^2;~m^2,m^2,4m^2)
= R_3(0,m^2,m^2,m^2;0,m^2,m^2).
\end{equation}
Integral on the right hand-side is in fact propagator type integral
with one massless line. Applying recurrence relations given in
Ref.~\cite{Tarasov:1997kx} this integral can be reduced to simpler 
integral:
\begin{eqnarray}
&& R_3(0,m^2,m^2,m^2;0,m^2,m^2)
\nonumber \\
&&=
\frac{1}{(i \pi^{d/2})^2}
\int \int \frac{d^dk_1 d^dk_2}
{k_1^2 (k_2^2-m^2)[(k_1-k_2)^2-m^2]^2[(k_1+q_1)^2-m^2]}
\nonumber \\
&&=\frac{2}{3(d-3)} J_{111}^{(d-2)}(m^2),
\end{eqnarray}
where
\begin{equation}
J_{111}^{(d)}(q^2) =\frac{1}{(i\pi^{d/2})^2}
\int \int \frac{d^dk_1 d^dk_2}{(k_1^2-m^2)[(k_1-k_2)^2-m^2]
[(k_2-q)^2-m^2]}.
\end{equation} 
%with $(a)_n\equiv\Gamma(a+n)/\Gamma(a)$.
At $q^2=m^2$,  the  result for 
$J_{111}^{(d)}(q^2)$ is known \cite{Broadhurst:1993mw}:
\begin{eqnarray}
&& \frac{m^{6-2d}}{8} (d-4)^2(d-2)(d-3) 
J_{111}^{(d)}(m^2)
\nonumber \\
&&~~~~~~~=\Fh32\Fu{1,3-d,\frac12}{ \frac{5-d}{2},\frac{d}{2}}
+(d-3)\Fh32\Fu{1,\frac{4-d}{2},\frac{d-1}{2}}{\frac32,d-1},
\end{eqnarray}
and it can be used for the $\varepsilon$ expansion of $R$ and $R_3$.
As was already mentioned at $q_1^2=q_2^2=m^2$, $q_3^2=4m^2$ 
integrals on the right hand - side of Eq.(\ref{fe_for_R}) correspond to propagator
type integrals. Analytic result for $R$ reads
\begin{eqnarray}
&&(d-3)R(m^2,m^2,m^2,m^2,m^2,m^2,4m^2)
%\Df{16}{3}m^2 J_{111}^{(d-2)}(m^2) + \left(T_1^{(d-2)}\right)^2
\nonumber \\
&&~~~~~~~~~~~=\frac{1}{8m^4(d-4)}
\left[ 2m^2(3d-8)(d-3) J_{111}^{(d)}-3(d-2)^2 \left( T_1^{(d)}\right)^2 \right],
\end{eqnarray}
where
\begin{equation}
T_1^{(d)}= -{\Gamma\left(1-\frac{d}{2}\right)}(m^2)^{(d-2)/2}.
\end{equation}                     
We checked that several first terms in the $\varepsilon=(4-d)/2$ expansion 
of $R$ and $R_3$ are in agreement with results of \cite{Kniehl:2008dt}.
The main profit from functional equations for $R$ and
$R_3$ comes from the fact that vertex integrals were expressed in
terms of simpler, propagator type integrals.
%\section{Auxiliary parameters and parametric representation}
%%%%%%%%%%%%%%%%%%%%%%%%%%%%%%%%%%%%%%%%%%%%%%%%%%%%%%%%%%%%%%%%
\section{Deriving functional equation by deforming  propagators}
%%%%%%%%%%%%%%%%%%%%%%%%%%%%%%%%%%%%%%%%%%%%%%%%%%%%%%%%%%%%%%
%%%%%%%%%%%%%%%%%%%%%%%%%%%%%%%%%%%%%%%%%%%%%%%%%%%%%%%%%%%%%%

The method described in the previous section does not work
for deriving functional equations for all kinds of Feynman integrals. 
For example, we did not found functional equation for the two-loop 
vacuum type integral given in Figure 4.
\begin{figure}[ht]
\begin{center}
\includegraphics[scale=0.6]{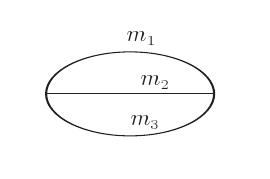}
\end{center}
\caption{Two-loop  vacuum type diagram }
\end{figure}
\vspace*{0.7cm}

In this section we shall describe another method that 
extends the class of integrals for which we can obtain functional
equations. The method is based on  transformation of
functional equations for some auxiliary integrals depending on arbitrary
parameters into functional equations for integrals of interest.
Such functional equations will be derived from algebraic relations
for `deformed propagators' which will be defined in the next section. 
These auxiliary integrals will be transformed into $\alpha$ parametric
representation. In general characteristic polynomials of
these integrals in $\alpha$ parametric representation  differ
from those for the investigated integral.  Functional equation
for the integral of interest can be obtained in  case
when it will be possible to map characteristic
polynomials of auxiliary integrals with `deformed propagators' 
to  characteristic polynomials of this integral. 
Such a mapping will be performed   by rescaling $\alpha$ parameters and
appropriate choice of arbitrary `deforming parameters'.

%%%%%%%%%%%%%%%%%%%%%%%%%%%%%%%%%%%%%%%%%%%%%%%%%%%%%%%%%%%%%%

\subsection{Algebraic relations for products of deformed propagators}

%%%%%%%%%%%%%%%%%%%%%%%%%%%%%%%%%%%%%%%%%%%%%%%%%%%%%%%%%%%%%%

In the previous section to derive  functional
equation we added to our consideration a propagator
with combination of external momenta taken with arbitrary
scalar coefficient. Now we consider generalization of this
method.

To find functional equation for $L$-loop Feynman 
integral depending on $E$- external momenta we  start from the 
relation of the form
\begin{equation}
\prod_{r=1}^{n}\frac{1}{D_r} = \frac{1}{D_{n+1}}
\sum_{r=1}^n x_r {\prod_{\substack{j=1
\\ 
j \neq r} }^n} \left(\frac{1}{D_j}\right),
\label{X_parameters}
\end{equation}
where  $D_j$ is defined as:
\begin{equation}
D_j=Q_j^2 - m_j^2+i\epsilon,
\label{deformed_prop}
\end{equation} 
with
\begin{equation}
Q_j=\sum_{l=1}^L a^{(j)}_{jl} k_l + \sum_{l=1}^{E+1} b^{(j)}_{jl} p_l,
\label{deformed_momentum}
\end{equation}
and $a^{(j)}_{jl}$, $b^{(j)}_{jl}$ for the time being are arbitrary 
scalar parameters.  Some of these parameters as well as
$x_r$ will be fixed from the equation (\ref{X_parameters}).
Another part of these parameters will be fixed from the 
requirement that the product of propagators in 
(\ref{X_parameters}) should correspond to the
integrand of the integral with the considered  topology. 
We would like to remark that instead of  deformation of propagators 
proposed in Eqs.~(\ref{deformed_prop}),(\ref{deformed_momentum}) 
one can use other deformations. For example, all terms in denominators
of propagators can be taken with arbitrary scalar coefficients:
\begin{equation}
Q_j^2=\sum_{l,r=1}^L a^{(j)}_{rl}  (k_r\cdot k_l)
+\sum_{l=1}^L \sum_{r=1}^{E+1} b^{(j)}_{rl} (p_r\cdot k_l)
+ \sum_{r,l=1}^{E+1} c^{(j)}_{rl} (p_r\cdot p_l).
\label{deformation2}
\end{equation}

To establish algebraic relation (\ref{X_parameters})
we put all terms  over  a common denominator and then
equate coefficients in front scalar products depending on integration momenta.
Solving obtained system of equations gives some restrictions
on the scalar parameters.
%from relation (\ref{X_parameters}).

In general integrals obtained by integrating products of `deformed
propagators'  will not correspond to  usual Feynman integrals. Further
restrictions on parameters should  be imposed  in order  to obtain relations
between integrals corresponding to Feynman integrals
coming from a realistic quantum field theory models.  

%%%%%%%%%%%%%%%%%%%%%%%%%%%%%%%%%%%%%%%%%%%%%%%%%%%%%%%%%%%%%%

\subsection{Functional equation for two-loop vacuum type integral with arbitrary masses}

%%%%%%%%%%%%%%%%%%%%%%%%%%%%%%%%%%%%%%%%%%%%%%%%%%%%%%%%%%%%%%

As an example, let us consider derivation of functional
equation for the two-loop vacuum type integral given in Figure 4:
\begin{equation}
J_0^{(d)}(m_1^2,m_2^2,m_3^2)
= \int\!\!\int\frac{\rd^dk_1\rd^dk_2}{(i \pi^{d/2})^2}
\frac{1}{(k_1^2-m_1^2)((k_1-k_2)^2-m_2^2)( k_2^2-m_3^2) }.
\label{J0def}
\end{equation} 
Analytic expression for this integral was presented in Ref
\cite{Davydychev:1992mt}.
Instead of this integral we will first consider an auxiliary integral
with integrand made from `deformed propagators' defined in Eqs.(\ref{deformed_prop}),
(\ref{deformed_momentum}):
\begin{equation}
\widetilde{J}_0^{(d)}(m_1^2,m_2^2,m_3^2)
=\int\!\!\int\frac{\rd^dk_1\rd^dk_2}{(i \pi^{d/2})^2}
\frac{1}{D_1 D_2 D_3 },
\label{J0deformed}
\end{equation}
where
\begin{eqnarray}
&& D_1 =(a_1 k_1+a_2k_2)^2 - m_1^2+i\epsilon,~~
   D_2 =(b_1 k_1+b_2k_2)^2 - m_2^2+i\epsilon,
\nonumber \\
&& D_3 =(h_1 k_1+h_2k_2)^2 - m_3^2+i\epsilon.~~
\label{D123_for2loop_bubble}
\end{eqnarray}
For the product of three deformed propagators one can try to find  an 
algebraic relation of the form:
\begin{equation}
\frac{1}{D_1 D_2 D_3} =   \frac{x_1}{D_4 D_2 D_3} + 
\frac{x_2}{D_1 D_4 D_3} + \frac{x_3}{D_1 D_2 D_4},
\label{2loop_3props}
\end{equation}
where $D_1$, $D_2$,$D_3$ are defined in Eq.(\ref{D123_for2loop_bubble})
and 
\begin{eqnarray}
D_4 =(r_1 k_1+r_2k_2)^2 - m_4^2.
\label{D4_for2loop_bubble}
\end{eqnarray}
Here $m_k$  are arbitrary masses, $x_k$,
 $a_j$, $b_i$, $h_s$, $r_l$  are undetermined parameters
and $k_1$, $k_2$  will be integration momenta. 

Putting  in Eq. (\ref{2loop_3props}) all over a common 
denominator and equating to zero coefficients in front of different
products of  $k_1^2$, $k_2^2$ and $k_1k_2$  leads to the system of
equations:
\begin{eqnarray}
&&r_1^2   - x_1 a_1^2   - x_2 b_1^2   - x_3 h_1^2 = 0,~~~
  r_1 r_2 - x_1 a_1 a_2 - x_2 b_1 b_2 - x_3 h_1 h_2 = 0,
\nonumber \\
&&
r_2^2  - x_1 a_2^2  - x_2 b_2^2  - x_3 h_2^2=0,
~~~~m_4^2-x_1 m_1^2-x_2 m_2^2-x_3 m_3^2=0.
\label{equs_4_j0}
\end{eqnarray}
%%%%%%%%%%%%%%%%%%%%%%%%%%%%%%%%%%%%%%%%%%%%%%%%%%%%%%%%%%
Solving this system for $r_1$, $x_1$, $x_2$, $x_3$ we have:
\begin{eqnarray}
&&r_1 = r_2 \lambda ,  
\nonumber 
\\
&&A x_1 = r_2^2 (h_1 h_2 m_2^2-b_2 b_1 m_3^2)
+b_2h_2(b_1h_2-b_2h_1)m_4^2
-r_2^2 (h_2^2 m_2^2-m_3^2 b_2^2) ~\lambda,
\nonumber 
\\
&&A x_2 = -r_2^2 (m_1^2 h_2 h_1-a_2 a_1 m_3^2)
-a_2h_2(a_1h_2-a_2h_1)m_4^2
 +r_2^2 (h_2^2 m_1^2-m_3^2a_2^2) ~\lambda, 
\nonumber 
\\
&& A x_3 = r_2^2 (m_1^2 b_2 b_1-  a_2 a_1 m_2^2)
+a_2b_2(a_1b_2-a_2b_1)m_4^2
-r_2^2(b_2^2 m_1^2-m_2^2a_2^2) ~\lambda,
\label{solution_4terms_mm4nonzero}
\end{eqnarray}
where $\lambda$ is a root of the quadratic equation
\begin{equation}
A \lambda^2+B\lambda+C=0,
\end{equation}
and
\begin{eqnarray}
&&A = 
b_2 h_2(b_1h_2-h_1b_2)m_1^2
+a_2h_2(h_1a_2-a_1h_2)m_2^2
+a_2b_2(a_1b_2-b_1a_2)m_3^2,
\nonumber 
\\      
&&
B = (h_1b_2-b_1h_2)(b_1h_2+h_1b_2)m_1^2
+(a_1h_2-h_1a_2)(a_1h_2+h_1a_2)m_2^2
\nonumber \\
&&~~~~~~~~~~~~~~~~~~~~~~~~~~~~~~~~~~~~~~~
+(b_1a_2-a_1b_2)(a_1b_2+b_1a_2)m_3^2,
\nonumber \\       
&&
C =b_1h_1(b_1h_2-h_1b_2) m_1^2
  +a_1h_1(a_2h_1-a_1h_2) m_2^2
  +a_1b_1(a_1b_2-b_1a_2) m_3^2
\nonumber \\
&&~~~~~~~
  +(b_2 h_1-b_1 h_2)(a_1 h_2-a_2h_1)(a_1b_2-a_2b_1)m_4^2.
\label{coefsABCnonzero_mm4}
\end{eqnarray}
%%%%%%%%%%%%%%%%%%%%%%%%%%%%%%%%%%%%%%%%%%%%%%%%%%%%%%%%%%
In order to obtain functional equation for the  integral
$J_0^{(d)}$, we integrate first both sides of the Eq.~(\ref{2loop_3props})
with respect to $k_1$, $k_2$ and then convert these integrals
into the $\alpha$-parametric representation.
%We apply standard methods (see for example, \cite{BogolyubovShirkov}).
Transforming all propagators into a parametric form
\begin{equation}
\frac{1}{(k^2-m^2+i\epsilon)^{\nu}}
 = \frac{i^{-\nu}}{ \Gamma(\nu)}\int_0^{\infty}
 d\alpha ~\alpha^{\nu-1} \exp\left[i\alpha(k^2-m^2+i\epsilon)\right],
\end{equation}
and using the $d$- dimensional Gaussian integration  formula
\begin{equation}
\int d^dk \exp \left[i(a k^2+ 2(pk))\right] =i
 \left( \frac{\pi}{i a} \right)^{\frac{d}{2}}
 \exp \left[ -\frac{ip^2}{a} \right] ,
\end{equation}
we can easily evaluate the integrals over loop momenta.
The final result is:
\begin{equation}
\widetilde{J}_0^{(d)}(m_1^2,m_2^2,m_3^2)\!=  i^{1-d}
~\prod^{3}_{j=1} 
 \int \limits_0^{\infty} \! \ldots \! \int \limits_0^{\infty}
\frac{d \alpha_j }
     { [ \widetilde{D}(\alpha) ]^{\frac{d}{2}}}
  \exp \left[-i \sum_{l=1}^{3}\alpha_l(m_l^2\!-\!i\epsilon)
                \right],
\label{repres}
\end{equation}
where the polynomial
\begin{equation}
\widetilde{D} = 
(a_1 b_2-a_2 b_1)^2\alpha_1 \alpha_2
+(a_1 h_2-a_2 h_1)^2\alpha_1 \alpha_3
+(b_1 h_2-b_2 h_1)^2\alpha_2 \alpha_3,
\end{equation}
differs from the appropriate $D$ polynomial  of the two-loop vacuum integral
defined in Eq.(\ref{J0def})
\begin{equation}
D = 
\alpha_1 \alpha_2
+\alpha_1 \alpha_3
+\alpha_2 \alpha_3.
\end{equation}
Changing in Eq. (\ref{repres}) integration variables 
%By rescaling $
\begin{equation}
\alpha_j \rightarrow \alpha_j \theta_j^2
\label{alf_rescaling}
\end{equation}
with
\begin{equation}
\theta_1 = b_1h_2-b_2h_1, ~~~
\theta_2 = a_1h_2-a_2h_1, ~~~
\theta_3 = a_1b_2-a_2b_1,
\label{theta123}
\end{equation}
leads to the relation:
\begin{equation}
\widetilde{D} = (b_1h_2-b_2h_1)^2(a_1h_2-a_2h_1)^2(a_1b_2-a_2b_1)^2
\left(\alpha_1 \alpha_2
+\alpha_1 \alpha_3
+\alpha_2 \alpha_3\right).
\end{equation}
Therefore  integral $\widetilde{J}_0^{(d)}$ with deformed 
propagators is proportional to ${J}_0^{(d)}$ with modified arguments:
%%%%%%%%%%%%%%%%%%%%%%%%%%%%%%%%%%%%%%%%%%%%%%%%%%%%%%%%%%%%%%%%%%%%%%%
\begin{eqnarray}
&&\int \int \frac{d^dk_1~d^dk_2}{(i\pi^{d/2})^2}
\frac{1}{D_1D_2D_3}
=i^{1-d}
[\theta_1^2 \theta_2^2 \theta_3^2]^{\frac{2-d}{2}}
\int \limits_0^{\infty}\int \limits_0^{\infty}\int 
\limits_0^{\infty}
\frac{d\alpha_1 d\alpha_2 d\alpha_3 }{D^{\frac{d}{2}}}
\exp \left[-i {\cal{M}}
                \right]
\nonumber \\
&&~~~~~~=[\theta_1^2 \theta_2^2 \theta_3^2]^{\frac{2-d}{2}}
J^{(d)}_0(\theta_1^2 m_1^2,\theta_2^2 m_2^2,\theta_3^2m_3^2)
\label{LHS_in_params2}
\end{eqnarray}
where
\begin{equation}
{ \cal{M}}= \alpha_1 \theta_1^2 m_1^2
    +\alpha_2 \theta_2^2 m_2^2
    +\alpha_3 \theta_3^2 m_3^2.
\end{equation}
With the aid of Eq.~(\ref{LHS_in_params2}) relation  (\ref{2loop_3props})
integrated with respect to $k_1$, $k_2$ can be written  as a combination 
of integrals  $J^{(d)}_0$ with different arguments:
%%%%%%%%%%%%%%%%%%%%%%%%%%%%%%%%%%%%%%%%%%%%%%%%%%%%%%%%%
\begin{eqnarray}
&&
[\theta_1^2 \theta_2^2 \theta_3^2]^{\frac{2-d}{2}}
J^{(d)}_0(\theta_1^2 m_1^2,\theta_2^2 m_2^2,\theta_3^2 m_3^2)
\nonumber \\
&&~~~~~ =x_1[\theta_1^2 \theta_6^2 \theta_4^2]^{\frac{2-d}{2}}
J^{(d)}_0(\theta_1^2 m_4^2,\theta_6^2 m_2^2 ,\theta_4^2 m_3^2)
\nonumber \\
%&& \nonumber
%\\
&&~~~~~+x_2[\theta_2^2 \theta_5^2 \theta_6^2]^{\frac{2-d}{2}}
J^{(d)}_0(\theta_6^2 m_1^2,\theta_2^2 m_4^2,\theta_5^2 m_3^2)
\nonumber \\
&&~~~~~+x_3[\theta_3^2 \theta_4^2 \theta_5^2]^{\frac{2-d}{2}}
J^{(d)}_0(\theta_4^2 m_1^2,\theta_5^2 m_2^2,\theta_3^2 m_4^2),
\label{func_equ_j0c}
\end{eqnarray}
%%%%%%%%%%%%%%%%%%%%%%%%%%%%%%%%%%%%%%%%%%%%%%%%%%%%%%%%%
where $\theta_1$,$\theta_2$,$\theta_3$ are defined in (\ref{theta123}) and
\begin{equation}
\theta_4 = r_1b_2-r_2b_1,~~~~~
\theta_5 = r_1a_2-r_2a_1,~~~~~
\theta_6 = r_1h_2-r_2h_1. 
\end{equation}
Now we consider relation (\ref{func_equ_j0c}) as an
equation for integrals in momentum representation.
By rescaling integration variables $k_1$, $k_2$ in the integral 
on the  left hand - side 
\begin{equation}
k_1= (\theta_1\theta_2\theta_3)^{\frac12}\widetilde{k}_1,~~~~~
k_2= (\theta_1\theta_2\theta_3)^{\frac12}\widetilde{k}_2,
\label{scaling_momenta}
\end{equation}
and performing analogous changes for the integrals on the right hand -
side we obtain the relation
%%%%%%%%%%%%%%%%%%%%%%%%%%%%%%%%%%%%%%%%%%%%%%%%%%%%%%%%%%%%%%%%%%%% vstavka% 
\begin{eqnarray}
\label{posle_mom_scale_mm4nz}
&&\frac{1}{{\theta_1 \theta_2 \theta_3}}
J_0^{(d)}\left( \frac{\theta_1}{\theta_2\theta_3} m_1^2,
{\frac{\theta_2}{\theta_1\theta_3}} m_2^2,
{\frac{\theta_3}{\theta_1\theta_2}} m_3^2\right)
\nonumber \\
&&~~~~~
=\frac{x_1}{{\theta_1 \theta_4\theta_6}} 
J_0^{(d)}\left({\frac{\theta_1}{\theta_4 \theta_6}}m_4^2,
         {\frac{\theta_6}{\theta_1 \theta_4}}m_2^2,
         {\frac{\theta_4}{\theta_1 \theta_6}}m_3^2\right)
\nonumber 
\\
&& \nonumber 
\\
&&~~~~+\frac{x_2}{{\theta_2\theta_5 \theta_6}}
J_0^{(d)}\left( {\frac{\theta_6}{\theta_2 \theta_5}}~m_1^2,
 {\frac{\theta_2}{\theta_6 \theta_5}}~m_4^2,
 {\frac{\theta_5}{\theta_2 \theta_6}}~m_3^2\right)
\nonumber \\
&&~~~~
+\frac{x_3}{{\theta_3 \theta_4 \theta_5}}
J_0^{(d)}\left({\frac{\theta_4}{\theta_3 \theta_5}}~m_1^2,
         {\frac{\theta_5}{\theta_3 \theta_4}}~m_2^2,
         {\frac{\theta_3}{\theta_5 \theta_4}}~m_4^2
\right).
\end{eqnarray}
In terms of redefined masses $M_1$, $M_2$, $M_3$ related to original
masses $m_1$, $m_2$, $m_3$ as
\begin{equation}
m_1^2= {\frac{\theta_2 \theta_3}{\theta_1}} M_1^2,~~~
m_2^2= {\frac{\theta_1 \theta_3}{\theta_2}} M_2^2,~~~
m_3^2= {\frac{\theta_1 \theta_2}{\theta_3}} M_3^2,
\label{chnge_masses}
\end{equation}
equation (\ref{posle_mom_scale_mm4nz}) reads
\begin{eqnarray}
&&J_0^{(d)}(M_1^2,M_2^2,M_3^2)
= {\frac{\theta_2\theta_3}
  {\widetilde{\theta}_4 \widetilde{\theta}_6}}
\widetilde{x}_1
J_0^{(d)}\left({\frac{\theta_1}{\widetilde{\theta}_4 \widetilde{\theta}_6  } } m_4^2,
{\frac{\theta_3 \widetilde{\theta}_6}
{\theta_2 \widetilde{\theta_4}}}~M_2^2,
{\frac{\theta_2 \widetilde{\theta}_4}
{\theta_3 \widetilde{\theta_6}}}~M_3^2
\right)
\nonumber 
\\
&&~~~~~~~~~~+{\frac{\theta_1\theta_3}{\widetilde{\theta}_5
\widetilde{\theta}_6}}~\widetilde{x}_2
J_0^{(d)}\left({\frac{\theta_3 \widetilde{\theta}_6}
{\theta_1\widetilde{\theta}_5}}M_1^2,
{\frac{\theta_2}{\widetilde{\theta}_5 \widetilde{\theta}_6  } } m_4^2,
{\frac{\theta_1\widetilde{\theta}_5}{\theta_3 \widetilde{\theta}_6} }
~M_3^2\right)
\nonumber 
\\
&&~~~~~~~~~~+{\frac{\theta_1\theta_2}{\widetilde{\theta}_4
\widetilde{\theta}_5}}~\widetilde{x}_3
J_0^{(d)}\left({\frac{\theta_2 \widetilde{\theta}_4}
{\theta_1\widetilde{\theta}_5}}M_1^2,
{\frac{\theta_1\widetilde{\theta}_5}{\theta_2 \widetilde{\theta}_4} }
~M_2^2,
{\frac{\theta_3}{\widetilde{\theta}_4 \widetilde{\theta}_5  } } m_4^2
\right),
\label{equ_with_redefined_masses_mm4nz}
\end{eqnarray}
where $\widetilde{x}_i$ , $\widetilde{\theta}_j$ are defined
from Eqs.~(\ref{solution_4terms_mm4nonzero}),(\ref{coefsABCnonzero_mm4})
with redefined masses.
After simplifications Eq.~(\ref{equ_with_redefined_masses_mm4nz}) 
takes a simpler form:
 \begin{eqnarray}
&&J_0^{(d)}(M^2_1,M^2_2,M^2_3)
\nonumber \\
&&=
J_0^{(d)}\left( \rho_1+\rho_2+M_1^2-M_2^2-M_3^2,\rho_1,\rho_2\right)
\nonumber \\
&&
+J_0^{(d)}\left(\rho_3, M_3^2-\rho_2, \rho_3-\rho_2 -M_1^2+M_2^2\right)
\nonumber \\
&&
-J_0^{(d)}\left(\rho_3-M_1^2,\rho_1-M_2^2, \rho_1+\rho_3 -M_3^2\right),
\label{equ_with_redefined_masses2}
\end{eqnarray}
where
\begin{eqnarray}
&&\rho_1=
\frac{
(M_3^2+M_2^2-M_1^2)\theta_1 \theta_2 \theta_3
-2b_2h_2\theta_2 \theta_3 m_4^2
-\sqrt{\delta_2} }
{2\theta_1 \theta_2 \theta_3M_2^2-2 b_2^2\theta_2^2m_4^2}
~M_2^2
\nonumber \\
&&\rho_2=\frac{(M_3^2+M_2^2-M_1^2)\theta_1 \theta_2 \theta_3
-2b_2h_2\theta_2\theta_3 m_4^2 +\sqrt{\delta_2}}
{2\theta_1 \theta_2 \theta_3M_3^2
-2h_2^2\theta_3^2m_4^2}~M_3^2
\nonumber \\
&&\rho_3=
\frac{(M_1^2-M_2^2+M_3^2)
\theta_1 \theta_2 \theta_3
+2a_2h_2\theta_1 \theta_3 m_4^2+\sqrt{\delta_2}}
{2\theta_1 \theta_2 \theta_3 M_1^2
-2a_2^2\theta_1^2 m_4^2}~M_1^2,
\end{eqnarray}
with
\begin{eqnarray}
&&\delta_2=\theta_1^2\theta_2^2\theta_3^2 \Delta_2 +4 A
\theta_1\theta_2\theta_3 m_4^2,
\nonumber \\
&&
\Delta_2 = M_1^4+M_2^4+M_3^4-2M_1^2M_2^2
-2M_1^2M_3^2-2M_2^2M_3^2.
\end{eqnarray}
The coefficinets $A$, $B$ and $C$ can be expressed in terms
of $M_i$, $\theta_j$:
\begin{eqnarray}
&&A=b_2h_2 \theta_2 \theta_3 M_1^2
-a_2h_2\theta_1 \theta_3 M_2^2+a_2b_2\theta_1 \theta_2M_3^2,
\nonumber \\
&&
B=
-(b_1h_2+b_2h_1)\theta_2\theta_3 M_1^2
+(a_1h_2+a_2h_1)\theta_1\theta_3 M_2^2
-(a_1b_2+a_2b_1)\theta_1\theta_2M_3^2,
\nonumber \\
&&
C=b_1h_1\theta_2 \theta_3 M_1^2-\theta_1\theta_3 a_1h_1M_2^2
+a_1b_1\theta_1 \theta_2 M_3^2
- \theta_1 \theta_2 \theta_3m_4^2.
\end{eqnarray}
%As one can see from (\ref{r02i3inro1})  
One can easily observe that due to relations
\begin{eqnarray}
&& \rho_1\rho_2=M_2^2M_3^3,
\nonumber \\
&&\rho_1 \rho_3 = M_1^2\rho_1+M_2^2\rho_3,
\nonumber \\
&&\rho_2\rho_3= M_3^2 \rho_3-M_1^2M_3^2,
\label{r02i3inro1}
\end{eqnarray}
parameters $\rho_2$ and $\rho_3$ can be  expressed 
in terms of $\rho_1$  and
therefore all  the auxiliary parameters introduced in derivation
of the functional equation will be absorbed only in one parameter - $\rho_1$.

We would like to notice that  Eq.~(\ref{equ_with_redefined_masses2})
is valid for integrals but not for their integrands.
This is due to the fact that the factor in front of integral that 
comes from the scaling of $\alpha$ 
parameters  in parametric integral is not fully compensated by 
scaling momenta given in Eq.~(\ref{scaling_momenta}).

At $m_4=0$  the  dependence on all parameters $a_i$,$b_j$,$h_k$ in 
Eqs.~(\ref{equ_with_redefined_masses_mm4nz}),
(\ref{equ_with_redefined_masses2}) drops out and the integral
$J_0^{(d)}(M^2_1,M^2_2,M^2_3)$ reduces to a comination of simpler
integrals:
\begin{eqnarray}
&&J_0^{(d)}(M^2_1,M^2_2,M^2_3)
\nonumber \\
&&=
J_0^{(d)}\left(0,\frac{-M^2_1+M^2_2+M^2_3+\sqrt{\Delta_2}}{2},
\frac{-M^2_1+M^2_2+M^2_3-\sqrt{\Delta_2}}{2}\right)
\nonumber \\
&&
+J_0^{(d)}\left(\frac{M^2_1-M^2_2+M^2_3+\sqrt{\Delta_2}}{2},
0,\frac{M^2_1-M^2_2+M^2_3-\sqrt{\Delta_2}}{2}\right)
\nonumber \\
&&
-J_0^{(d)}\left(\frac{-M^2_1-M^2_2+M^2_3+\sqrt{\Delta_2}}{2},
\frac{-M^2_1-M^2_2+M^2_3-\sqrt{\Delta_2}}{2},0\right).
\label{J0_mm4_zero}
\end{eqnarray}

%%%%%%%%%%%%%%%%%%%%%%%%%%%%%%%%%%%%%%%%%%%%%%%%%%%vstavka %%%%%%%%%%%%%%%%

%Parameters $x_1$,$x_2$,$x_3$, and $r_1$ are given in Eq. (\ref{solution_4terms}). 
Analytic expression for the integral $J_0^{(d)}$ with one mass equal to zero is 
known \cite{Davydychev:1992mt}. Under assumption that  
$\left|m_3^2- m_2^2\right| \le \left| m_3^2\right|$ it reads
\begin{equation}
J_0^{(d)}(0,m_2^2,m_3^2)=
\frac{\pi^2(2-d)\left( m_2^2m_3^2\right)^{\frac{d-2}{2}}}
{2(d-3)m_3^2 \left(\Gamma\left(\frac{d}{2}\right)
\sin \frac{\pi d}{2}\right)^2 }~
\Fh21\FmM{1,2-\frac{d}{2}}{4-d}.
\label{J0_hgf}
\end{equation} 
From functional equation (\ref{J0_mm4_zero}) as a by-product one can
get a new hypergeometric representation for the one-loop massless vertex
type integral.
In Ref.~\cite{Davydychev:1995mq} an interesting relation between the dimensionally
 regularized  one-loop vertex type integral $I_3^{(d)}(m_1^2, m_2^2, m_3^2)$
and  the two-dimensional integral 
$J_0^{(6-d)}(m_1^2, m_2^2, m_3^2)$  was discovered 
\begin{equation}
I^{(d)}_3(s_{23},s_{13},s_{12})=
\frac{\Gamma\left(3-\frac{d}{2}\right)}
{\Gamma(d-3)}
(-s_{23}s_{13}s_{12})^{\frac{d-4}{2}}  J_0^{(6-d)}(s_{23},s_{13},s_{12}).
\end{equation}
Functional equation (\ref{J0_mm4_zero}) with $J^{(d)}_0(0,m_2^2,m_3^2)$
defined in Eq.~(\ref{J0_hgf}) provide us a new hypergeometric 
representation for the integral $I_3^{(d)}$ with massless propagators.
Formula for the one-loop massless vertex integral
in terms of other Gauss' hypergeometric functions 
is given in Ref.~\cite{Davydychev:1999mq}.

%%%%%%%%%%%%%%%%%%%%%%%%%%%%%%%%%%%%%%%%%%%%%%%%%%%%%%

%%%%%%%%%%%%%%%%%%%%%%%%%%%%%%%%%%%%%%%%%%%%%%

%\input{sect_conclusions}

\section{Conclusions}

Finally, we summarize what we have accomplished in this paper.

First of all, we formulated new methods for 
deriving functional equations for Feynman integrals.
These methods are rather simple and do not use any kind
of integration by parts techniques.

Second, it was shown that integrals with many
kinematic arguments can be reduced to a combination
of simpler integrals with fewer arguments.
In our future publications we are going to demonstrate 
that in some cases applying functional equations one can  reduce, the so-called,
master integrals to a combination of simpler integrals  from, what we 
would like to call,  a `universal' basis of integrals.

The method based on algebraic relations for `deformed propagators'
can be used not only for vacuum type of integrals but 
also for integrals depending on external momenta.
In the present paper we considered rather particular cases 
of functional equations. 
 The systematic investigation and classification of
the proposed functional equations requires application
of the methods of algebraic geometry and group theory.

At the present moment it is not quite clear whether 
functional equations derivable from recurrence relations can
be reproduced by the methods of algebraic relations between
products of propagators described in Section 3 and Section 5.

A detailed consideration of our functional 
equations and their application to the one-loop
integrals with four,  five and six external legs
as well as  to some two- and three- loop  Feynman 
integrals will  be presented in future publications.

\section{Acknowledgment}
This work was supported by the German Science Foundation (DFG) within
the Collaborative Research Center 676  
{\it Particle, Strings and the Early Universe: the Structure of Matter and Space-Time}.
%, project $\rm B4$.
I am thankful to O.L. Veretin for providing  results for integrals contributing 
to ortho-positronium lifetime described in Ref.\cite{Kniehl:2008dt}.

\end{document}